\documentclass{webofc}
\usepackage[varg]{txfonts}   
\providecommand{\sorthelp}[1]{}

\newcommand{\ngal}{{N_{\rm gal}}}
\newcommand{\Sx}{{S_{\rm X}}}
\newcommand{\Sxprime}{{S'_{\rm X}}}
\newcommand{\Sstd}{{S_{\rm std}}}

\def\lesssim{\mathrel{\hbox{\rlap{\hbox{\lower4pt\hbox{$\sim$}}}\hbox{$<$}}}}
\let\la=\lesssim  			
\def\gtrsim{\mathrel{\hbox{\rlap{\hbox{\lower4pt\hbox{$\sim$}}}\hbox{$>$}}}}

\def\ngal{N_{\rm gal}}
\def\Sx{S_{\rm X}}
\def\Sxprime{S'_{\rm X}}
\def\Sstd{S_{\rm std}}

\begin{document}
\title{Biases in the estimation of velocity dispersions and dynamical masses for galaxy clusters}
%
%

\author{\firstname{A.} \lastname{Ferragamo}\inst{1,2}\fnsep\thanks{\email{ferragamo_ext@iac.es}} \and 
	\firstname{J.~A.} \lastname{Rubi\~{n}o-Mart\'{\i}n}\inst{1,2} 
	\and
	\firstname{J.} \lastname{Betancort-Rijo}\inst{1,2} 
	\and
	\firstname{E.} \lastname{Munari}\inst{3,4} 
	\and
	\firstname{B.} \lastname{Sartoris}\inst{3,4} 
	\and
	\firstname{R.} \lastname{Barrena}\inst{1,2}
}

\institute{Instituto de Astrof\'{\i}sica de Canarias (IAC), C/ V\'{\i}a L\'actea s/n, E-38205 La Laguna, Tenerife, Spain
  \and Universidad de La Laguna, Departamento de Astrof\'{\i}sica, C/ Astrof\'{\i}sico Francisco S\'anchez s/n, E-38206 La Laguna, Tenerife, Spain  
  \and Dipartimento di Fisica, Sezione di Astronomia, Universit\`a di Trieste, Via Tiepolo 11, I-34143 Trieste, Italy
  \and INAF/Osservatorio Astronomico di Trieste, Via Tiepolo 11, I-34143 Trieste, Italy}

\abstract{%
Using a set of 73 numerically simulated galaxy clusters, we have characterised
  the statistical and physical biases for three velocity dispersion
  and mass estimators, namely biweight, gapper and standard deviation, in the
  small number of galaxies regime ($N_{\rm gal} \le 75$), both for the determination of the
  velocity dispersion and the dynamical mass of the clusters via the
  $\sigma$--$M$ relation. These results are used to define a new set of unbiased
  estimators, that are able to correct for those statistical biases. 
  By applying these new estimators to a subset of simulated observations, we show
  that they can retrieve bias-corrected values for both the mean velocity
  dispersion and the mean mass.
  }
\maketitle
\section{Introduction}
\label{intro}
Several authors have used the velocity dispersion mass proxy to study and
characterise scaling relations between SZ and dynamical mass \citep{ruel14,
  sifon16, amodeo17}. 
In order to have sample with enough statistical power, it is necessary to estimate the velocity dispersion for hundreds of galaxy clusters (GCs). Although this goal can be achieved through spectroscopic follow-up \cite[e.g.][]{nostro16, rafa18}, these studies are extremely expensive in terms of observational time and data reduction. 
For these reasons, it is extremely difficult to estimate radial velocities for more than few members (typically $\sim 20$) for each cluster target.

In this article we present our study of statistical and physical biases introduced in the estimation of velocity dispersion and dynamical mass.

\section{Statistical biases in Velocity Dispersion estimation}
\label{stats}
For our analysis we use a sample of 73 simulated
massive clusters selected from the simulations described in
\cite{munari13}. Our selected sample contains clusters with masses $M_{200} > 2 \times 10^{14} M_\odot$ and located at five redshifts between $0.12 \leq z \leq 0.82$.

There are several method to estimate mean and scale of a distribution. We focus our attention on the estimators that, in the last decades, became standard tools in GC analyses, namely biweight and gapper, compared with the standard deviation. A detailed description of these estimators can be found in \cite{beers90}. 
\begin{figure*}
\centering
\includegraphics[trim=0.4cm 0.4cm 0.4cm 0.4cm, clip, width=0.49\textwidth]{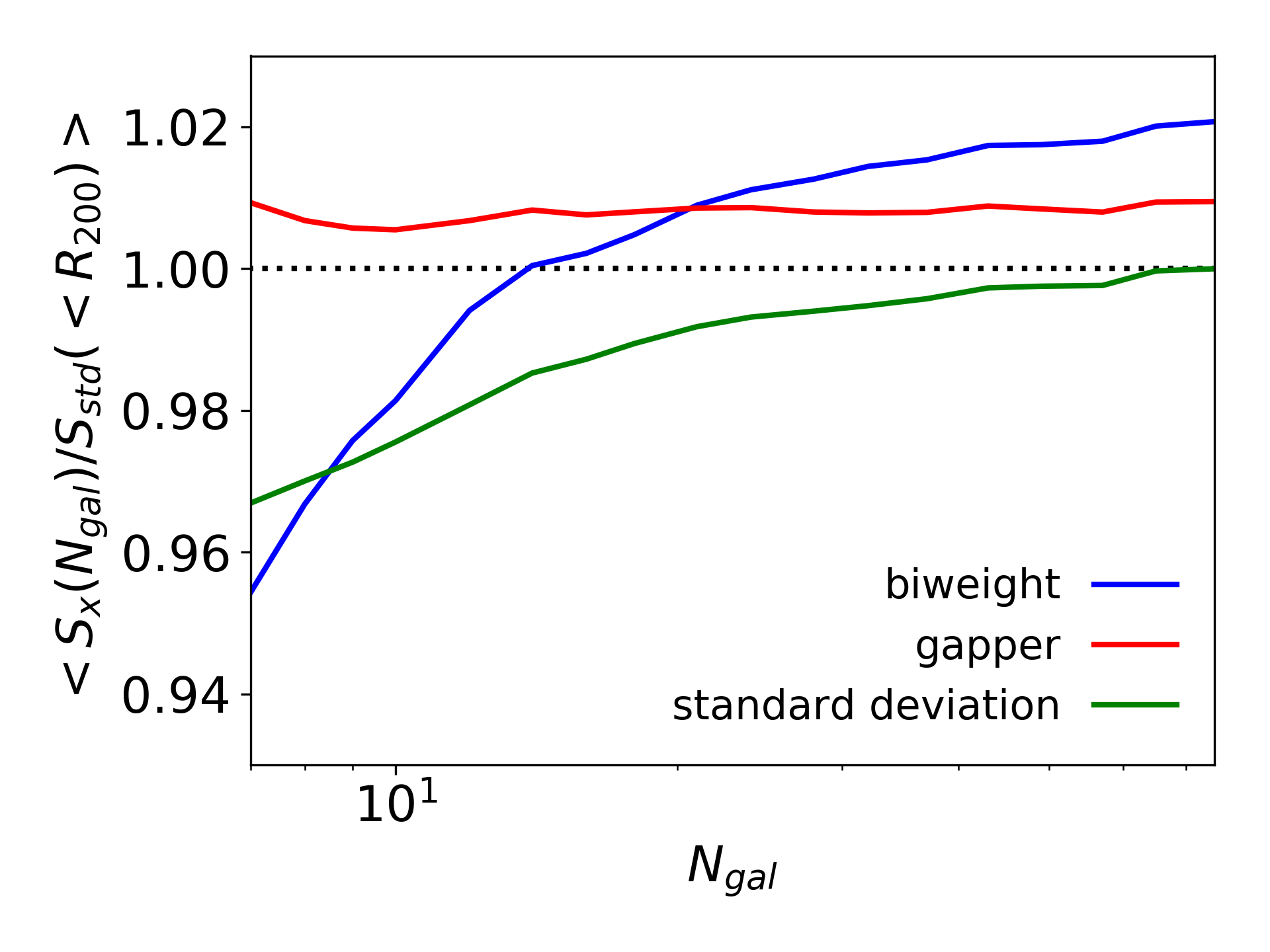}
\includegraphics[trim=0.4cm 0.4cm 0.4cm 0.4cm, clip, width=0.49\textwidth]{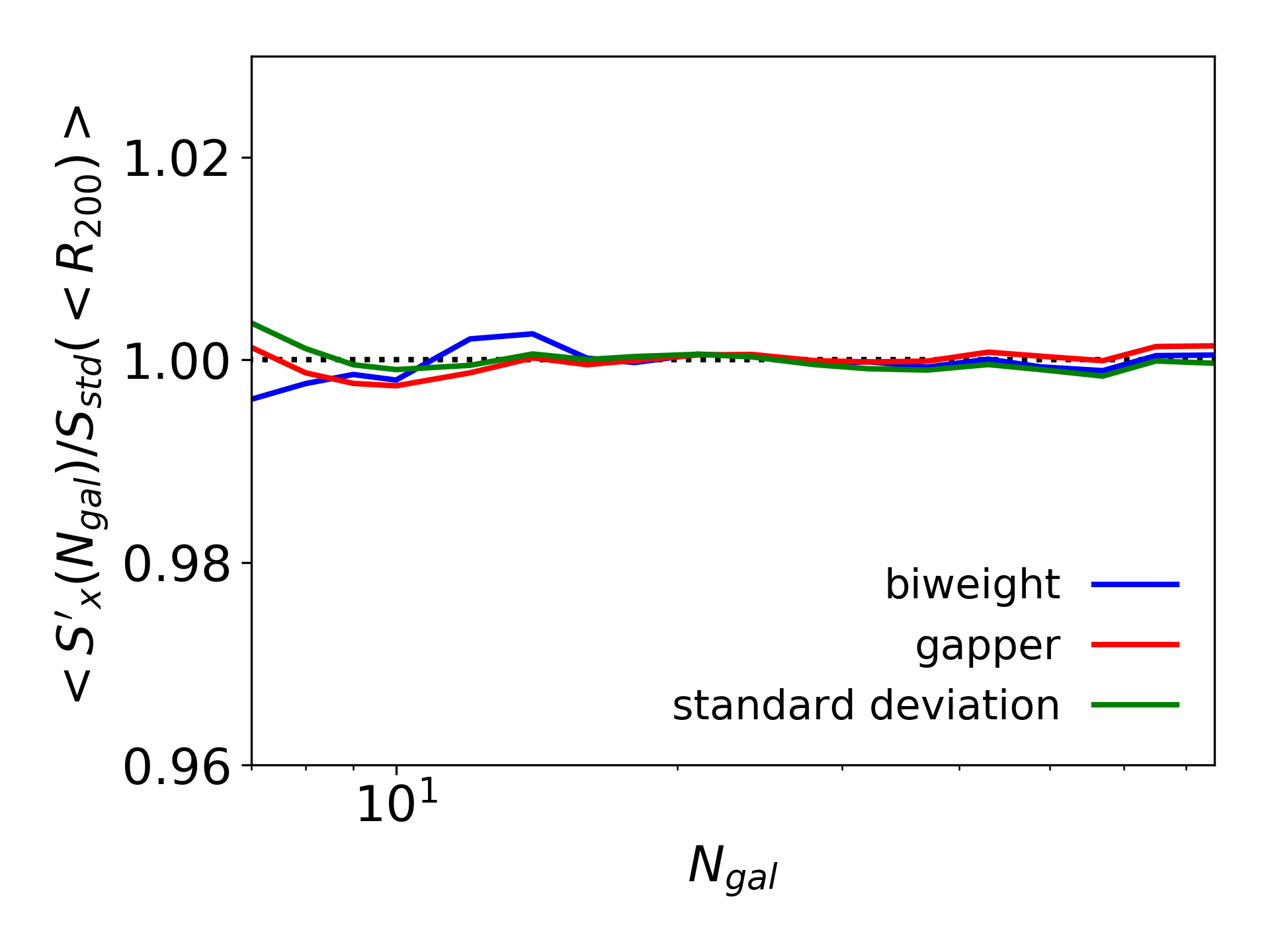}
\caption{\textit{Left panel:} mean velocity dispersion $\Sx / \Sstd(<R_{200})$ as a function of
  the number of galaxies $\ngal$. \textit{Right panel:} corrected velocity dispersion estimators $\Sxprime / \Sstd(<R_{200})$ as a function of $\ngal$. The dispersion $\Sx(\ngal)$ is
  calculated for the standard deviation (green line), biweight (blue line), and
  gapper (red line) estimators. }
\label{fig:n_sigma}       
\end{figure*}
In order to investigate bias and variance of the three scale estimators
$\Sx(\ngal)$, being X$=$ ``std'', ``bwt'' or ``gap'', for each one of the three cases, we have explored the low number of galaxies regime between $\ngal =8$ and $\ngal =75$.
We have generated 2500 configurations of randomly selecting
galaxies within the projected circle of radius $R_{200}$ and then we averaged the $73 \times 2250$ velocity dispersions (750 for each mayor axis).%

Left panel of Fig.~\ref{fig:n_sigma} shows the average $\Sx(\ngal)$ normalised with respect to $\Sstd(<R_{200})$, which represents the velocity dispersion of all the galaxies in the simulation within a circle of projected radius $R_{200}$, and calculated using the standard deviation estimator.
In the low-$\ngal$ regime, all estimators are biased showing very different behaviours.
The standard deviation estimator (green line) shows a dependence with the number of elements used for the estimation. However, this dependence can be theoretically predicted to be $1 -1/(4 (\ngal
-1))$.
The biweight (blue line) shows a stronger drop for $\ngal \leq 30$ underestimating the reference dispersion by up to $4$\,\% at $\ngal = 7$. Finally, the gapper (red line) shows an estimate of the velocity dispersion almost constantly biased at any $\ngal$. 

In order to construct an unbiased velocity dispersion estimators, $\Sxprime(\ngal)$, we use a parametrisation of the curves in left panel of Fig.~\ref{fig:n_sigma}: 
\begin{equation}
\label{eq:un_S}
\Sxprime(\ngal) \equiv \Sx(\ngal) \left(1 + \left(\left(\frac{D}{(\ngal -1)}\right)^\beta+B\right)\right).
\end{equation}

Table~\ref{table:param_un_sim} shows the best-fit values for the parameters $D$,
$\beta$ and $B$, for each one of the three estimators (biweight, gapper and
standard deviation).
\begin{table}
\footnotesize
\caption{Best-fit parameters to be used in the parametric function given in
  equation~\ref{eq:un_S}, describing the bias of the three
  estimators. See text for details.  }
\label{table:param_un_sim} 
\centering 
\begin{tabular}{c c c c} 
\hline\hline 
\noalign{\smallskip}
\, & $BWT$ & $GAP$ & $STD$ \\
\hline
\noalign{\smallskip}
$D$ & $0.72\pm0.03$ & $0$ & $0.25$\\ 
$B$ & $-0.0225\pm0.0002$ & $-0.0080\pm0.0002$ & $-0.0037\pm0.0003$\\
$\beta$ & $1.28\pm0.03$ & $ 1 $ & $ 1 $ \\
\noalign{\smallskip}
\hline 
\end{tabular}
\end{table}
Right panel of Fig.~\ref{fig:n_sigma} shows that the corrected estimators $\Sxprime(\ngal)$ are actually unbiased by construction.
%

\section{Biases by interlopers contamination}
\label{inter}
\begin{figure}
\begin{center}
\includegraphics[trim=0.5cm 0.5cm 0.5cm 0.5cm,clip,width=0.5\textwidth]{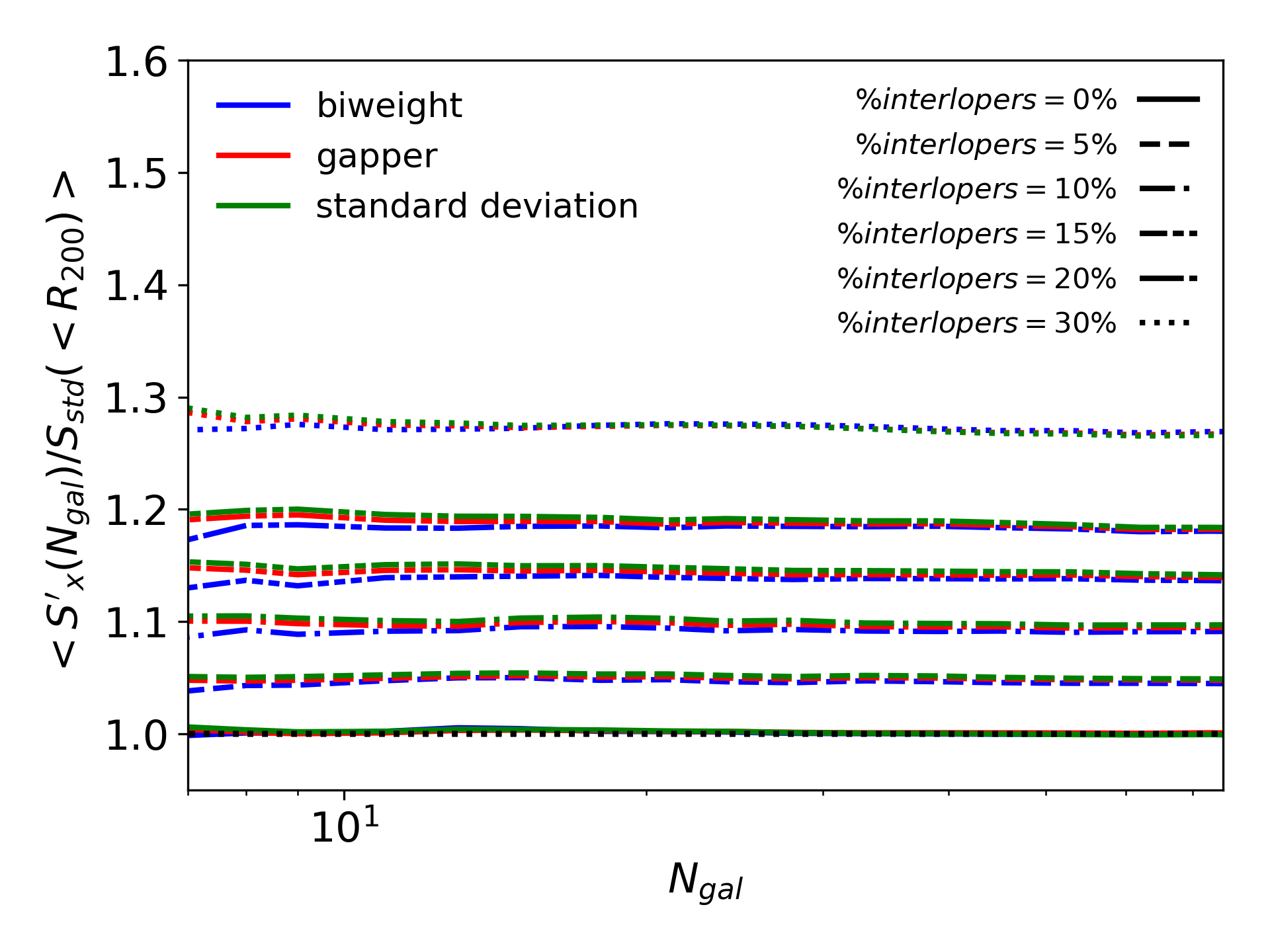}
\end{center}
\caption{Effect of interlopers on $\Sxprime/\Sstd(<R_{200})$, as a
 function of the number of galaxies. The velocity dispersion,
 $\Sxprime(\ngal)$, is first computed using a pure galaxie sample (solid
 lines). We also evaluated the response of biweight (blue lines), standard
 deviation (green lines), and gapper (red lines) using samples contaminated by
 $0\%$ (solid line), $5\%$ (dashed lines), $10\%$ (dot dashed lines), $15\%$ (three-dot-dashed
 lines), $20\%$ (two-dot-long-dashed lines) and $30\%$ (dotted
 lines) of interlopers at any $\ngal$. }
 \label{fig:interlopers}
\end{figure}

Any spectroscopic sample of cluster members is contaminated
by galaxies belonging to the large scale structure that surrounds the cluster itself. 
This population of `pseudo cluster members', called interlopers, modifies the velocity distribution and therefore
affects the estimation of velocity dispersion.
According to the definition of interlopers given in \cite{Pratt19}, one
must distinguish between two very different types of contaminants:
i. galaxies gravitationally bound to the clusters that are outside the virial sphere (according to the definition given in \cite{mamon10}), but that, due to projection effects appear within a smaller radius;
ii. background/foreground galaxies with similar redshifts to the cluster, but belonging to the large scale structure that surrounds the cluster itself.
Detailed study of these interlopers is beyond the scope of this work. Here, we illustrate the robustness of the three estimators in exam, by fixing the fraction of contaminants at any $\ngal$. Fig.~\ref{fig:interlopers} shows the results obtained for five
different fractions of interlopers: 5\% (dashed lines), 10\% (dot-dashed lines),
15\%(three-dot-dashed lines), 20\% (two-dot-long-dashed lines) and 30\% (dotted
lines). We note that the three estimators are similarly affected by interloper contamination. The effect of the interlopers contamination consists of a velocity dispersion overestimation which is as high as their relative fraction (up to 30\%) representing the most damaging source of biases in the velocity estimation estimation.

\section{Physical biases in Velocity Dispersion estimation}
\label{physi}

Observational strategies and technical limitations generally force us to observe only massive clusters members sampled in a fraction of $R_{200}$. We studied how these limitations might produce biased velocity dispersion estimates.


\subsection{Effects due to the selected fraction of massive galaxies}
\label{subsec:bias_mass}
In the ideal case we can observe any cluster member regardless of its brightness.
However, the telescope aperture limits the detection
magnitude and prevents us from detecting faint objects, for a fixed integration time.
Therefore, cluster samples contain only a fraction of the brightest galaxy members, which are also the most massive. 

In order to simulate this effect, we mimicked observational conditions by
selecting three percentages of all visible galaxies in the simulation,
i.e. 50\,\%, 33\,\%, and 25\,\%, by sorting the cluster members by mass and
dividing the sample in 2, 3, and 4 mass bins, starting from the most massive
object. For each case, we reproduce the procedure explained in Sect.~\ref{stats}.

\begin{figure*}
\centering
\includegraphics[trim=0.6cm 0.6cm 0.6cm 0.6cm, clip,width=0.49\textwidth]{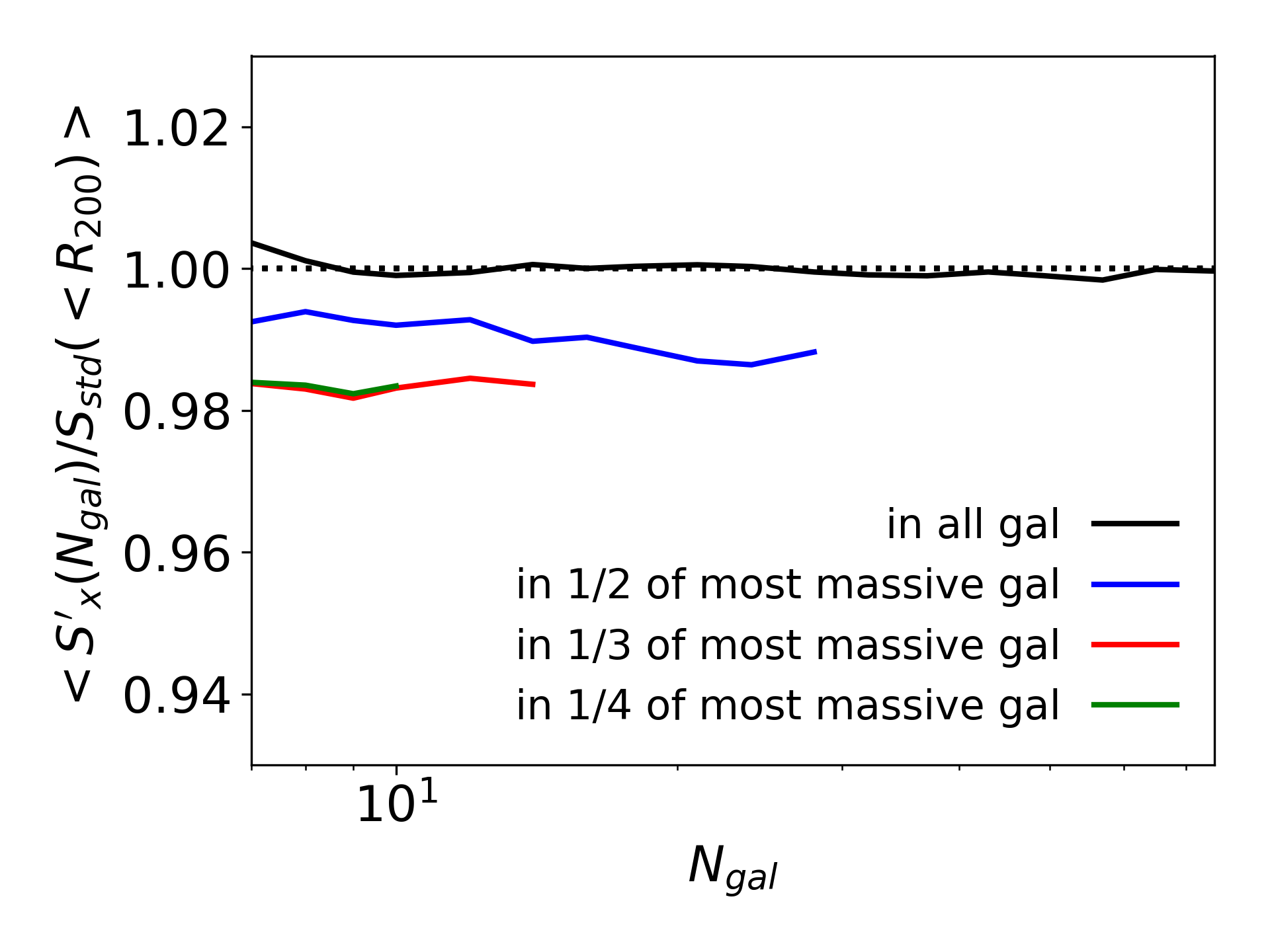}
\includegraphics[trim=0.6cm 0.6cm 0.6cm 0.6cm, clip,width=0.49\textwidth]{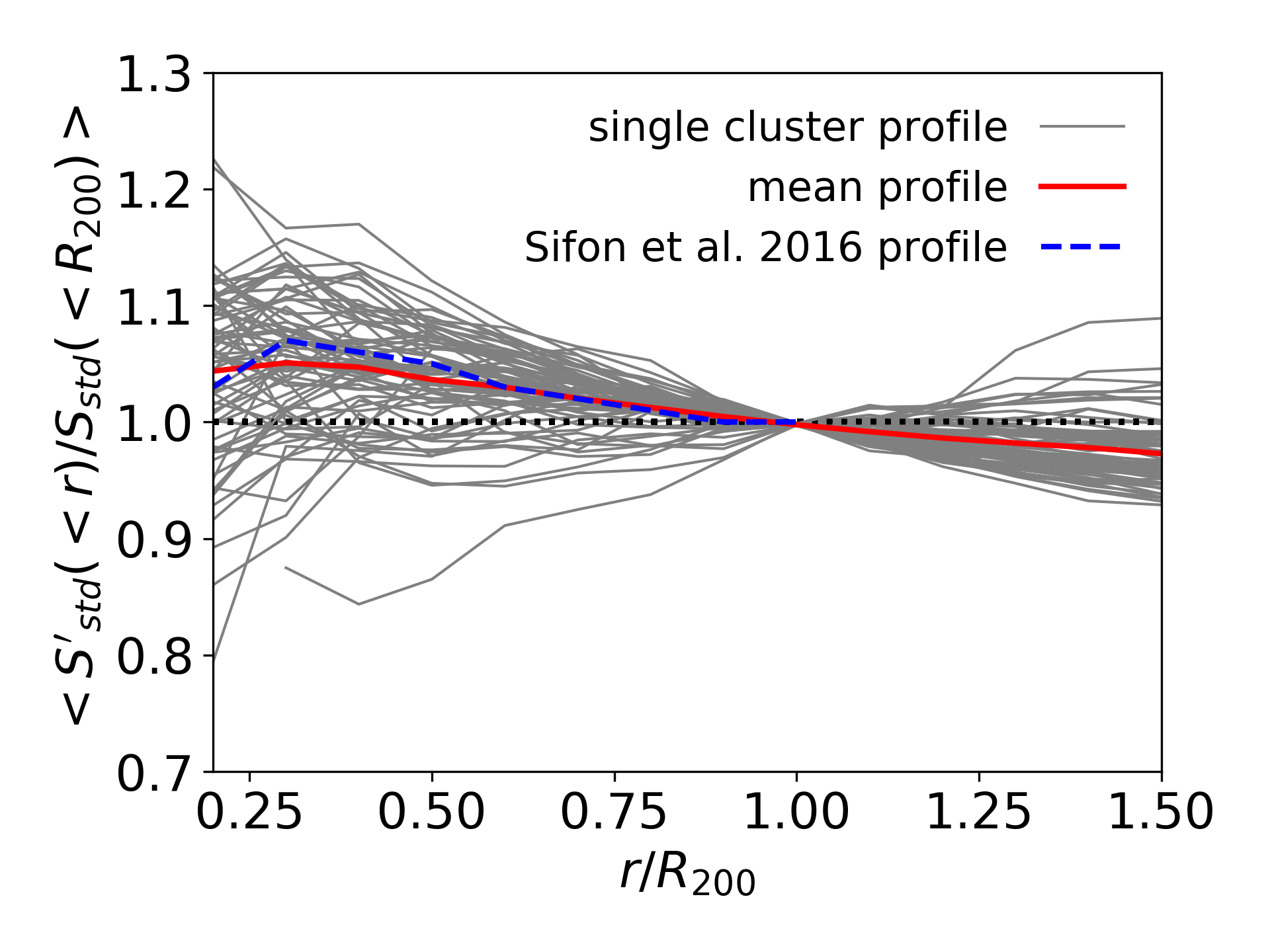}
\caption{\textit{Left panel:} Mean (bias) of $S'_{\rm std}(\ngal)/ \Sstd(<R_{200})$ as a
  function of the number of galaxies $\ngal$, calculated by choosing galaxies
  within $100\%$ (black solid line), $1/2$ (blue solid line), $1/3$ (red solid
  line), and $1/4$ (green solid line) of the complete cluster member
  samples. \textit{Right panel:} Average velocity dispersion profile within a given enclosed radius $r$,
  $<S'_{\rm std}(<r)/\Sstd(<R_{200})>$, normalised to $R_{200}$. The red line
  represents the mean at each radius of the individual 73 simulated GC profiles
  (grey lines). The dashed blue line represents the \cite{sifon16} profile, which is almost
  coincident with our derived profile.}
  \label{fig:n_sigma_ahtqo}
\end{figure*}

Fig.~\ref{fig:n_sigma_ahtqo} (left panel) shows $S'_{\rm
  std}(\ngal)/\Sstd(<R_{200})$ as function of $\ngal$ calculated with the
corrected standard deviation estimator, and using galaxies picked up from
100\,\% (black line), $1/2$ (blue line), $1/3$ (red line) and $1/4$ (green line)
of the complete cluster member samples. We see that the 
velocity dispersion is sensitive to the fraction of massive galaxies used to
estimate it, reaching a bias of almost 2\% using only the most massive fraction. However, it is almost insensitive to $\ngal$. For this reason, we can use the curves in the left panel of Fig.~\ref{fig:n_sigma_ahtqo} to correct this physical effect.

\subsection{Effect of aperture sub-sampling}
\label{subsec:bias3}
There are evidence in the literature that the velocity dispersion has a radial dependence \citep[e.g.,][]{mamon10, sifon16}. This implies that sampling galaxies in different fractions of the cluster's virial radius should lead to a biased velocity dispersion.
In order to investigate and quantify this effect, we averaged the 73 velocity dispersions
calculated using all galaxies inside a cylinder of variable radius $0.2\leq r/R_{200}
\leq 1.2$. Fig.~\ref{fig:n_sigma_ahtqo} (right panel) shows that the velocity dispersion is (on average) overestimated in region smaller than $R_{200}$ and slightly underestimated for $r>R_{200}$, in analogy with that recover by \cite{sifon16}.
\section{Statistical bias in the estimation of $M_{200}$}
\label{sec:s_mass_bias}


%
\begin{figure*}
\centering
\includegraphics[trim=0.5cm 0.5cm 0.5cm 0.5cm, clip,width=0.49\textwidth]{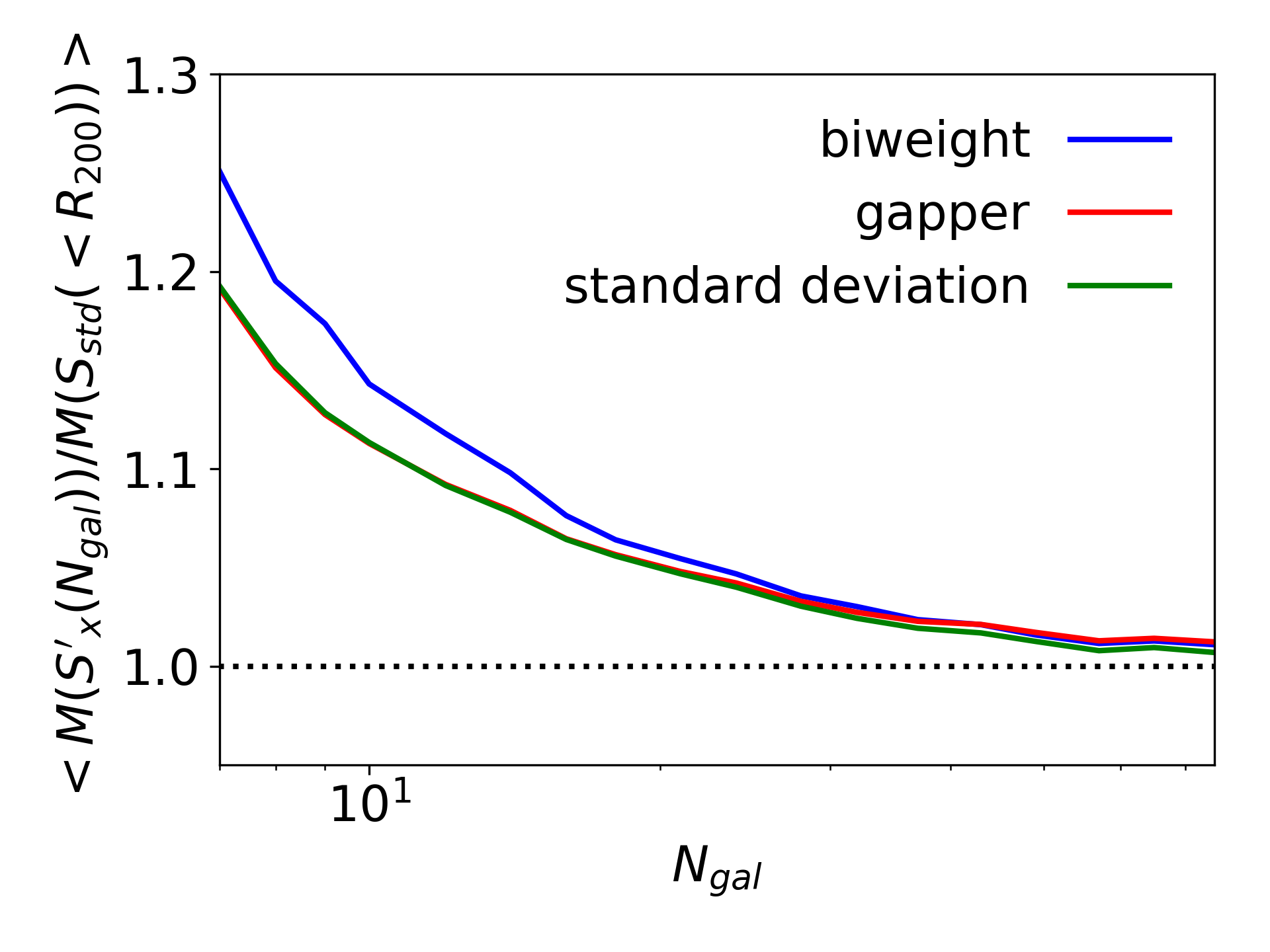}
\includegraphics[trim=0.5cm 0.5cm 0.5cm 0.5cm, clip,width=0.49\textwidth]{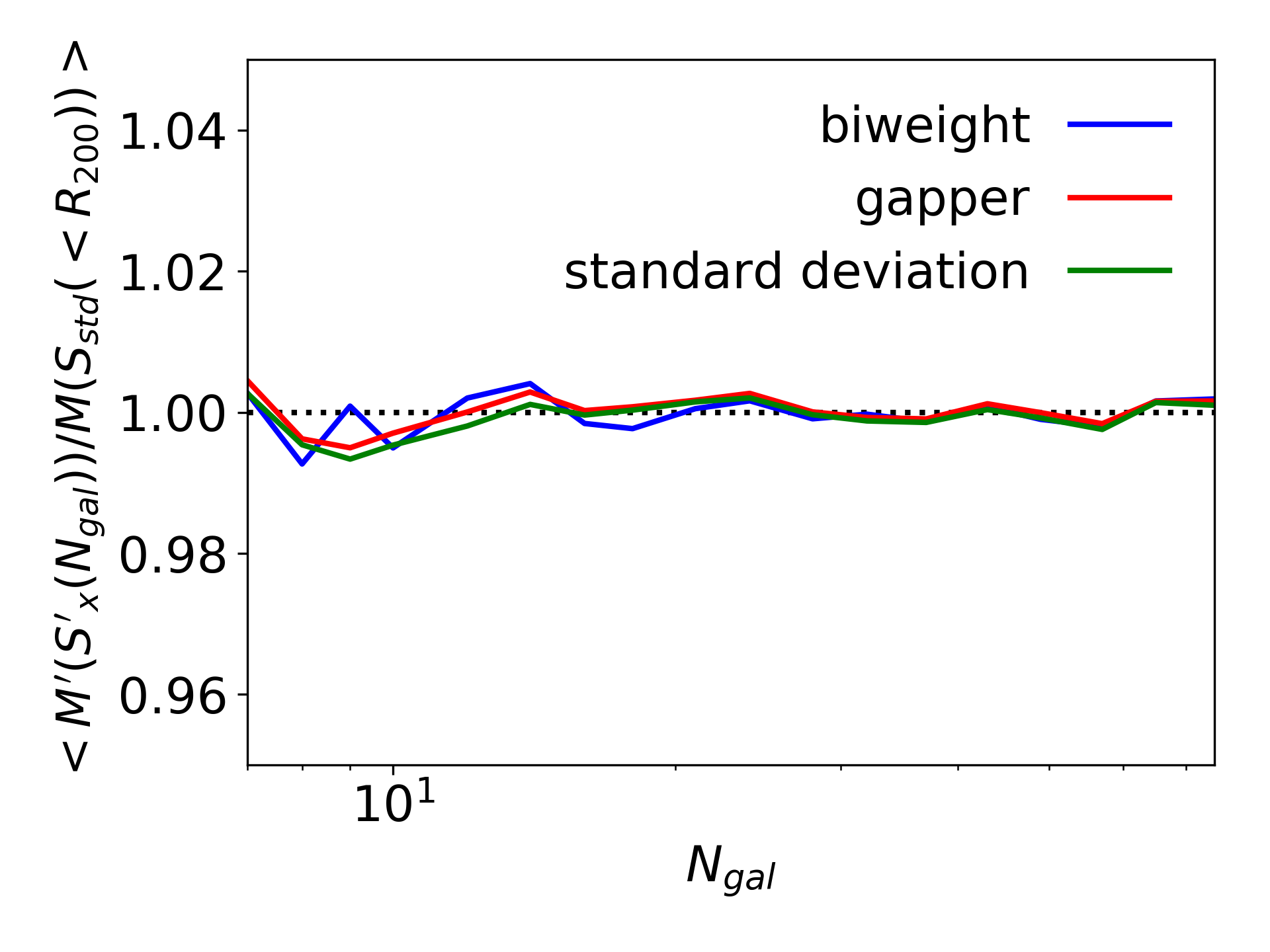}
\caption{\textit{Left panel:} mean of $M\left(\Sxprime(\ngal) \right)/M\left(S'_{\rm std}(<R_{200})\right)$, which represent the standard mass estimator, described in \cite{munari13}, applied to normal and unbiased velocity
  dispersion estimators, standard deviation (green), gapper (red), and biweight
  (blue). \textit{Right panel:} mean of the corrected mass estimator $M'\left(\Sxprime(\ngal) \right)/M\left(S'_{\rm std}(<R_{200})\right)$.}
  \label{fig:m_b_sim_f}
\end{figure*}

When we use velocity dispersion as a mass proxy we apply a scaling relation
$\sigma_{\rm 1D} - M$, it is a power law that has been previously calibrated either with simulations \cite{evrard08, saro13, munari13}. 
The left panel of Fig.~\ref{fig:m_b_sim_f} shows how applying the scaling relation $\sigma_{\rm 1D} - M$ to the unbiased velocity dispersion estimator $\Sxprime$ the resulting masses $\left< M(\Sxprime) /
M(\Sstd(<R_{200})) \right>$ are biased.

We can calculate analytically the functional form of the mass estimator as a function of the number of galaxies $\frac{1-2\alpha}{ 4\alpha^2 (\ngal-1)}$, with $A=1177.0$\,km\,s$^{-1}$ and $\alpha=0.364$ parameters of the \cite{munari13} scaling relation.

As in Sect.~\ref{stats}, we fitted a parametric form description of
the bias as a function of $\ngal$ by using three parameters ($E'$,$F'$ and
$\gamma'$, listed in Table~\ref{table:Fs'_par_sim}):
\begin{equation}
M'\left( \Sxprime(\ngal) \right) = M\left(\Sxprime(\ngal) \right)\left[\frac{1-E'\alpha}{(E'\alpha)^2(N_{\rm gal}-1)^{\gamma^\prime}}+F'\right]^{-1}.
\label{eq:M_M"_eq}
\end{equation}

\begin{table}
\footnotesize
\caption{Best-fit parameters for the function describing the bias in $\left<
  M(\Sxprime)/M(\Sstd(<R_{200})) \right>$ for simulated clusters, as described
  in equation~\ref{eq:M_M"_eq}.}
\label{table:Fs'_par_sim} 
\centering 
\begin{tabular}{c c c c} 
\hline\hline 
\noalign{\smallskip}
\, & $BWT$ & $GAP$ & $STD$ \\
\hline
\noalign{\smallskip}
$E'$ & $1.31\pm0.03$ & $1.50\pm0.03$    & $1.53\pm0.03$\\ 
$F'$ & $ 1 $                   & $ 1 $                      & $ 1 $\\
$\gamma^\prime$ & $1.24\pm0.03$ & $ 1.17\pm0.04 $  & $1.11\pm0.04 $ \\
\noalign{\smallskip}
\hline 
\end{tabular}
\end{table}
The right panel of Fig.~\ref{fig:m_b_sim_f} shows the bias of $M'(\Sxprime)$. The new mass estimator is actually unbiased by construction. 

Finally, in order to test the corrections described in previous sections, we have applied them to a set of mock cluster catalogues. To do this we have simulated a realistic observational strategy based on the observations described in \citep{nostro16,rafa18}. We generated $100$ mock samples out of the $73$ GCs object
simulated in this study. For each of these samples we calculated
the mean ratio between the estimated and the reference velocity dispersion of
each cluster, as well as the velocity dispersion calculated using $\Sxprime$ defined in eq.~\ref{eq:un_S} with the parameters in Table~\ref{table:param_un_sim}. Averaging over all the mock samples we obtained a biased mean velocity dispersion
\begin{equation} \label{eq:S_pre_corr}
\begin{split}
\left< \Sstd(\ngal,r)/S_{\rm bwt}(<R_{200}) \right> & = 0.96\pm0.02,\\
\left< S'_{\rm std}(N_{\rm gal},r)/S_{\rm bwt}(<R_{200})\right> & = 1.00\pm0.02.
\end{split}
\end{equation}

The normal $\Sstd$ estimator shows to be biased, whereas the $\Sxprime$ lead to an unbiased estimation of the velocity dispersion.

Using the bias corrected velocity dispersions, we calculated cluster masses,
$M(\Sxprime)$ and $M'\left( S'_{\rm std}\right)$, obtaining
\begin{equation} \label{eq:Fsp_post_corr}
\begin{split}
\left< M\left( S'_{\rm std}(N_{\rm gal},r)\right)/M\left(S_{\rm bwt}(<R_{200})\right)\right> & = 1.13\pm0.07,\\
\left< M'\left( S'_{\rm std}(N_{\rm gal},r)\right)/M'\left(S_{\rm bwt}(<R_{200})\right)\right> &= 1.00\pm0.06.
\end{split}
\end{equation}

As described above, these normal mass estimator is overestimated, while primed mass estimator, $M'$ is actually unbiased. 

\section{Conclusions}
\label{subsec:end}
We have used 73 simulated GCs from hydrodynamic simulations
including AGN feedback and star formation, in order to characterise the
statistical and physical biases in three velocity dispersion (and mass)
estimators frequently used in the literature: the biweight, the gapper, and the
standard deviation. We have focused our study in the (common) case of a low
number of galaxy members ($\ngal \la 75$).

We showed that each of these estimators (dispersion and mass) presents a statistical bias. Therefore, we propose bias corrected velocity dispersion ($\Sxprime$) and mass ($M'(\Sxprime)$) estimators.

We have tested the robustness of the new estimators against the contamination by interlopers. We found that the velocity dispersion estimators $\Sxprime$ are similarly affected by the
contamination for all the three cases in this low-$\ngal$ limit. 

We observed that the most likely sources of physical bias are i) the selection effect the fraction of massive galaxies used to estimate the velocity dispersion; and ii) the fraction of the virial radius explored. We showed that in the first case the bias is estimated to be around 2\,\% when considering only $1/4$ of the most massive galaxies.
Concerning the effect produced by the sampling aperture, we find a dispersion radial profile in
agreement with \cite{sifon16} results.


%


%
 \bibliography{bibliography_nightmare.bib}

\begin{thebibliography}{11}

\bibitem{ruel14}
J.~{Ruel}, G.~{Bazin}, M.~{Bayliss}, M.~{Brodwin}, R.J. {Foley}, B.~{Stalder},
  K.A. {Aird}, R.~{Armstrong}, M.L.N. {Ashby}, M.~{Bautz} et~al., \apj
  \textbf{792}, 45 (2014), \texttt{1311.4953}

\bibitem{sifon16}
C.~{Sif{\'o}n}, N.~{Battaglia}, M.~{Hasselfield}, F.~{Menanteau}, L.F.
  {Barrientos}, J.R. {Bond}, D.~{Crichton}, M.J. {Devlin}, R.~{D{\"u}nner},
  M.~{Hilton} et~al., \mnras \textbf{461}, 248 (2016), \texttt{1512.00910}

\bibitem{amodeo17}
S.~{Amodeo}, S.~{Mei}, S.A. {Stanford}, J.G. {Bartlett}, J.B. {Melin}, C.R.
  {Lawrence}, R.R. {Chary}, H.~{Shim}, F.~{Marleau}, D.~{Stern}, \apj
  \textbf{844}, 101 (2017), \texttt{1704.07891}

\bibitem{nostro16}
{Planck Collaboration Int. XXXVI}, \aap \textbf{586}, A139 (2016),
  \texttt{1504.04583}

\bibitem{rafa18}
R.~{Barrena}, A.~{Streblyanska}, A.~{Ferragamo}, J.A. {Rubino-Martin},
  A.~{Aguado-Barahona}, D.~{Tramonte}, R.T. {Genova-Santos}, A.~{Hempel},
  H.~{Lietzen}, N.~{Aghanim} et~al., ArXiv e-prints  (2018),
  \texttt{1803.05764}

\bibitem{munari13}
E.~{Munari}, A.~{Biviano}, S.~{Borgani}, G.~{Murante}, D.~{Fabjan}, \mnras
  \textbf{430}, 2638 (2013), \texttt{1301.1682}

\bibitem{beers90}
T.C. {Beers}, K.~{Flynn}, K.~{Gebhardt}, \aj \textbf{100}, 32 (1990)

\bibitem{Pratt19}
G.W. {Pratt}, M.~{Arnaud}, A.~{Biviano}, D.~{Eckert}, S.~{Ettori}, D.~{Nagai},
  N.~{Okabe}, T.H. {Reiprich}, \ssr \textbf{215}, 25 (2019),
  \texttt{1902.10837}

\bibitem{mamon10}
G.A. {Mamon}, A.~{Biviano}, G.~{Murante}, \aap \textbf{520}, A30 (2010),
  \texttt{1003.0033}

\bibitem{evrard08}
A.E. {Evrard}, J.~{Bialek}, M.~{Busha}, M.~{White}, S.~{Habib}, K.~{Heitmann},
  M.~{Warren}, E.~{Rasia}, G.~{Tormen}, L.~{Moscardini} et~al., \apj
  \textbf{672}, 122-137 (2008), \texttt{astro-ph/0702241}

\bibitem{saro13}
A.~{Saro}, J.J. {Mohr}, G.~{Bazin}, K.~{Dolag}, \apj \textbf{772}, 47 (2013),
  \texttt{1203.5708}

\end{thebibliography}
%
%
%
%

\end{document}